# Time-Domain Hybrid PAM for Data-Rate and Distance Adaptive UWOC System

T. Kodama[1], M. Aizat[1], F. Kobori[1], T. Kimura[2], and Y. Inoue[3], M. Jinno[1]
1. Faculty of Engineering and Design, Kagawa University, Takamatsu 761-0396 Japan
2. Faculty of Science and Engineering, Doshisha University, Kyotanabe 610-0394 Japan
3. Department of Information and Communication Technology, Graduate School of Engineering, Osaka University, Suita 565-0821 Japan

**Abstract:** The challenge for next-generation underwater optical wireless communication systems is to develop optical transceivers that can operate with low power consumption by maximizing the transmission capacity according to the transmission distance between transmitters and receivers. This study proposes an underwater wireless optical communication (UWOC) system using an optical transceiver with an optimum transmission rate for the deep sea with near-pure water properties. As a method for actualizing an optical transceiver with an optimum transmission rate in a UWOC system, time-domain hybrid pulse amplitude modulation (PAM) (TDHP) using a transmission rate and distance-adaptive intensity modulation/direct detection optical transceiver is considered. In the TDHP method, variable transmission capacity is actualized while changing the generation ratio of two intensity-modulated signals with different noise immunities in the time domain. Three different color laser diodes (LDs), red, blue, and green are used in an underwater channel transmission transceiver that comprises the LD and a photodiode. The maximum transmission distance while changing the incidence of PAM 2 and PAM 4 signals that calibrate the TDHP in a pure transmission line and how the maximum transmission distance changes when the optical transmitter/receiver spatial optical system is altered from the optimum conditions are clarified based on numerical calculation and simulation. To the best knowledge of the authors, there is no other research on data-rate and distance adaptive UWOC system that applies the TDHP signal with power optimization between two modulation formats.

## 1. Introduction

Underwater environments represent the last unexplored frontier. As technology has advanced, many human activities underwater such as seafloor research has progressed, which has led to many discoveries. Among these are the abundant natural resources such as petroleum and natural gas on the seabed. Furthermore, by observing underwater trenches and the movements of ocean plates researchers could predict tsunami and earthquake activity. Over time, seafloor related research and developments have progressed in areas such as disaster prevention of tsunami and earthquakes, the off-shore oil industry, diver-submarine communications [1], and underwater environment monitoring systems. However, exploring the deep seafloor has been deemed too dangerous for humans as the human body cannot withstand the high pressures associated with the depths. As the complexity and precision of robots has increased and they have proved reliable, they have become entrusted to explore underwater sites. Thus, a reliable means of communications must be established between the operator and machine to give the device instructions. An underwater communications system is valuable since it is the only way to transmit information from the sensors.

In general, there are two communication methods: wired and wireless. Wired communications are the most stable because there is no attenuation effect from the surrounding environment, so signal attenuation is kept at a minimum. However, due to the restrictions on wired connections, the machine cannot move freely. For use over a broad range in wireless communications, the standard communication method is to use electromagnetic waves [2]. However, there is a big hurdle to implement this type of communication system underwater. Since electromagnetic waves are attenuated easily underwater, alternatives such as acoustic waves have become popular due to their low attenuation and long-range transmission capability. The main problem with acoustic waves is the narrow bandwidth, which leads to high latency and high-power consumption. There are efforts to use other types of waves such as very low-frequency radio waves used by naval submarines to penetrate the depths to the seafloor. The low frequency causes natural background noise to increase and requires more power to overcome the noise. Due to this, acoustic waves are not reliable for high-speed underwater transmission. These problems have hindered large-scale use of underwater devices. Due to the few options for reliable underwater transmission, underwater-based machine development has stagnated for many years.

The one of major problem for the current onshore earthquake prediction system is the delay in earthquake prediction. Recent studies have shown that earthquake prediction techniques using seafloor observations are more accurate than the traditional onshore earthquake prediction technique that is still in use today [3]. The seafloor observation method can be developed into a higher-performance system by arranging many underwater Internet-of-Things (IoT) sensors on the seafloor. Underwater wireless optical communication (UWOC) system has been expected to establish a high-capacity underwater communication system [4,5]. However, to operate this type of system efficiently, an UWOC system must be constructed in advance [6]. Therefore, research and development on optical transceivers for UWOC systems are required [7].

Fig. 1 shows the overall structure of the UWOC system. The previously proposed UWOC system consists of three main components [4]: the access network, core network, and wired-wireless integration. This investigation focuses on the transceiver specifications for a short-reach underwater channel. For the transmitter, we use a light-emitting diode (LED) [8] and laser diode (LD) [9].

Signal-to-noise ratio (SNR) adaptive time-domain hybrid pulse amplitude modulation (PAM) (TDHP) [10-13] that uses two-level PAM (PAM 2) [14] and four-level PAM (PAM 4) signals together [15, 16] at the transceiver was proposed for optical fiber communication and indoor wireless optical communication systems to address the changes in SNR. The TDHP method is advantageous because it maximizes the transmission capacity by flexibly setting two types of PAM-$M$ signal ratios to the received SNR, which changes according to the water turbidity and transmission distance. The TDHP method can also change only the operation mode in digital signal processing (DSP), and does not require optical and radio frequency (RF) analog components. Since the operation of the TDHP method can be controlled using only software, it is easy to implement and is system scalable. In UWOC transmission, TDHP uses a single carrier that can be received by lowering the signal rate while maintaining the received SNR symbol rate, which is difficult with a pure PAM 4 signal.

As opposed to other research, we use the TDHP method to transmit data in an underwater channel using the UWOC system. Past research mainly used the orthogonal frequency-division multiplexing (OFDM) method for underwater and indoor transmission [17, 18]. The TDHP signal has a lower peak-to-average power ratio (PAPR) than the OFDM signal. The input voltage received by the transmitter driver amplifier is limited, so the average power transmitted from the LD is considerable.

In this paper, we consider a TDHP-based optical transceiver comprising an LD that is included in the core and access networks to maximize the transmission rate and distance corresponding to the received SNR. The SNR changes according to the transmission distance and water turbidity when the transmission distance between each wireless terminal is different in the UWOC system.

As a contribution to the research field of Gigabit-class high-speed underwater optical wireless communication system, the transmission capacity and transmission distance are maximized according to the transmission environment by simply changing two parameters of our proposed TDHP by software control in the transceiver. The novel point is to introduce a flexible modulation for a single carrier system used in the transceiver. The required transmission capacity can be controlled by changing the generation ratio of PAM2 and PAM4 format, the first parameters. By changing the average power of PAM2 and PAM4, which are the second parameters, the maximum transmission distance can be achieved with respect to the set transmission capacity. In this modulation method for pure water channel condition, the effect of optimizing the optical system on typical three visible light RGB colors is also verified by numerical analysis and Monte-Carlo simulation for the first time.

This paper is organized as follows. In Section II, the principle behind submarine optical wireless communication is explained. In Section III, we make the simulation model of TDHP and the calculation model of the maximum transmission distance. In Section IV, we show two results: noise tolerance of TDHP signals and transmission characteristics in seafloor transmission channel. In Section V, we discuss the case of misalignment and fine adjustments to the optical system setup.

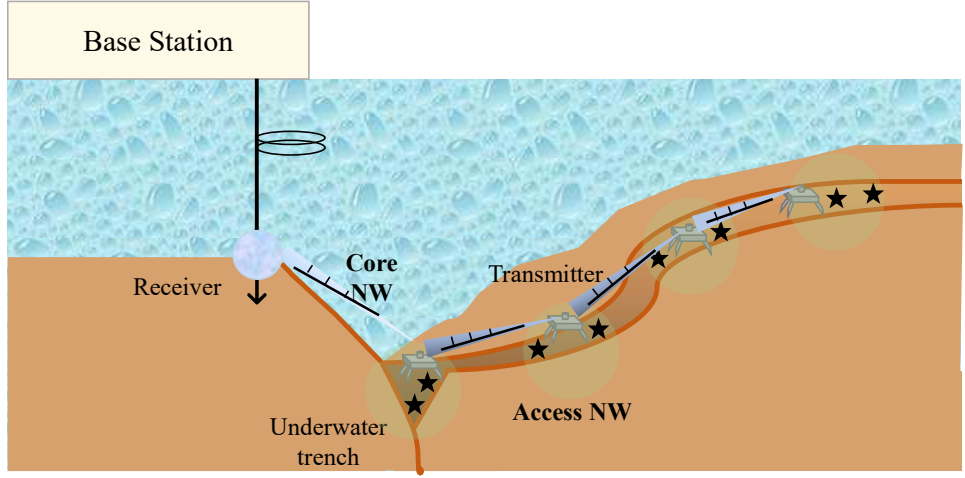

FIGURE 1. Overall structure of UWOC system.

## 2. Principle behind seafloor optical wireless communications

### A. PROPERTIES OF SEAFLOOR
The transmitter transmits the TDHP signal via an underwater channel, so the underwater channel properties and their effect on the TDHP signal should be considered.

The Beer-Lambert (BL) law concerns the absorption of radiant energy by an absorbing medium and in this case, water particles are used to evaluate the optical properties of the seafloor [19]. One of the main factors affecting signal deterioration in the UWOC system is signal attenuation due to the absorption and scattering effects underwater. The power delivered can be calculated by using the BL law as

$$P_d = P_0 \exp(-K_T d) \qquad (1)$$

where $P_0$ is the transmitted power, $K_T$ is the beam attenuation coefficient, and $d$ is the distance between the transmitter and the receiver. Beam attenuation coefficient $K_T$ can be calculated using

$$K_T = K_A + K_S \qquad (2)$$

where $K_A$ is the absorption rate coefficient and $K_S$ is the scattering coefficient. Absorption coefficient $K_A$ is further calculated using

$$K_A = K_{A0} + Ta \qquad (3)$$

Here, $K_{A0}$ is the non-temperature-dependent absorption rate coefficient, $T$ is the water temperature, and $a$ is the temperature coefficient. This equation shows that the signal absorption effect depends on the temperature of the underwater channel.

Based on (1), we show that the BL law only considers attenuation due to two factors: absorption and scattering. It does not assume any collection of scattered light that contributes to the received power. The reason for this is that some of the scattered light will be collected by the receiver [20]. The scattered light will affect lower received power caused by the significant scattering effect of highly turbid water.

The attenuation rate depends on the turbidity of the water, i.e., the murkier the water, the higher the attenuation rate. Table I gives the chlorophyll concentration, Gelbstoff concentration, plankton concentration, and the optimum wavelength of three water types: pure water, seawater, and dirty water [21]. Chlorophyll is a green pigment found in algae that allows it to absorb energy from sunlight.

Gelbstoff is dissolved yellow organic matter that causes the seawater to appear green, yellow-green, or brown [22]. The concentrations of chlorophyll, Gelbstoff, and plankton reflect the turbidity of the water. The properties of the water types are investigated to reflect the common real-world underwater channel of the UWOC system. Based on Table I, the chlorophyll concentration, Gelbstoff concentration, and plankton concentration are the lowest for pure water, followed by seawater, and dirty water. The optimum wavelength is the highest for dirty water and seawater at 520-570 nm, while pure water has the lowest optimum wavelength at 450-500 nm. These results suggest that the higher the water turbidity, the higher the optimum wavelength of the LD/LED needed to overcome the attenuation effect.

To understand further the optical properties of the seafloor, we reviewed previous research on the optical properties of deep-sea sites in the Mediterranean conducted by the Neutrino Mediterranean Observatory (NEMO) project in 1999 using a neutrino telescope deployed on the seafloor off the coast of Italy [23]. The absorption coefficient and the attenuation coefficient in the seas around Italy at various water depths were calculated for three years. This study concluded that pure water conditions follow the Smith and Baker model of pure water conditions [24]. Also, the passage of time has little effect on the optical properties of the seafloor. Thus, long-term usage of the UWOC system is viable.

TABLE I
WATER CHARACTERISTICS

| Water Type | Chlorophyll Concentration | Gelbstoff Concentration | Plankton Concentration | Optimum Wavelength |
|---|---|---|---|---|
| Pure water Seafloor water | Low | Low | Low | 450-500nm (blue-green) |
| Shallow sea water | High | High | High | 520-570 nm (yellow-green) |
| Dirty water | Very high | Very high | Very high | 520-570 nm (yellow-green) |

**B. SPATIAL OPTICAL CONFIGURATION**

This section describes the functions, properties, and factors affecting the effectiveness of the LD and photodetector (PD) for the UWOC system. Generally, the Spatial Optical System (SOS) factors are transmission power $P_t$, beam attenuation coefficient $K$, and transmission length $d$. Fig. 2 shows that the transceiver layout consists of an LD and PD [25].

An LD is a module that converts an electrical signal into an optical intensity signal to be transmitted via the underwater channel. The factors affecting the LD are the half-angle of the transmitter beamwidth, $\theta$, and the Noise Equivalent Power (NEP) [26]. The NEP is a standard metric that quantifies the PD sensitivity or the power generated by a noise source.

A PD is a module that receives and converts the transmitted optical intensity signal from an LD into an electric signal again. The factors affecting the PD are receiver aperture diameter $D$, the receiver field-of-view (FOV), and angle misalignment between the line-of-sight (LOS) and the optical axis of the receiver.

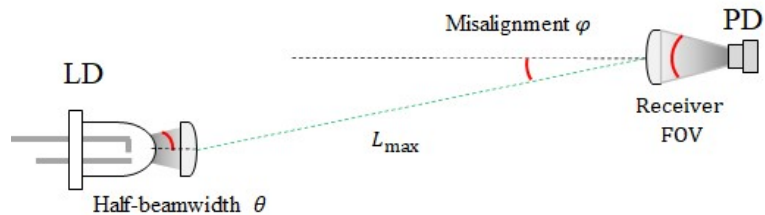

FIGURE 2. Structure of optical components.

**C. TIME-DOMAIN HYBRID PAM SIGNAL**

The TDHP signal comprises conventional PAM 2 and PAM 4 signals [27]. Table II gives the SNR tolerance and the number of bits per symbol for pure water PAM 2 (Pure PAM 2), TDHP, and pure PAM 4 signals. The pure PAM 2 signal achieves a high SNR tolerance [28], but it has a lower data rate of 1 bit per symbol. In contrast, the pure PAM 4 signal has a low SNR tolerance and a higher data rate of 2 bits per symbol [29, 30]. On the other hand, the TDHP signal SNR tolerance and the number of bits per symbol is between that for pure PAM 2 and pure PAM 4.

Parameters of PAM 4 generation ratio $p$ and power control factor $q$ are used to optimize the TDHP signal. The equation below defines the effects of the $p$ and $q$ overall performance of the TDHP bit error rate (BER).

$$\text{BER}_{PAM2} = \frac{1}{2}\text{erfc}\left(\sqrt{\frac{1}{2}\text{SNR}(1-q)}\right) \quad (4)$$

$$\text{BER}_{PAM4} = \frac{3}{8}\text{erfc}\left(\sqrt{\frac{1}{14}\text{SNR}(1+q)}\right) \quad (5)$$

$$\text{BER}_{TDHP} = p\text{BER}_{PAM4} + (1-p)\text{BER}_{PAM2} \quad (6)$$

where $\text{BER}_{PAM2}$, $\text{BER}_{PAM4}$, and $\text{BER}_{TDHP}$ are the BERs for the PAM 2, PAM 4, and TDHP signals, respectively. In (6), $p$ is used to control the PAM 2 and PAM 4 occurrence ratio inside TDHP so that the TDHP has an optimized data rate [31]. The $p$ parameter controls the $\text{BER}_{PAM2}$ and $\text{BER}_{PAM4}$ inside the transmission distance of pure water. By increasing $p$, the PAM 4 BER will increase while that for PAM 2 will decrease. Since the PAM 4 signal has a higher bit rate, the overall rate will increase. On the other hand, $q$ is used to control both the signal power of PAM 2 and PAM 4. The $q$ parameter controls the SNR for PAM 2 and PAM 4. As shown in (4) and (5), by increasing $q$, the PAM 4 power ratio will increase while that for PAM 2 will decrease. This leads to better performance for PAM 4 at the cost of worse performance for PAM 2. However, the overall TDHP performance will be better. The $p$ and $q$ parameters must be adjusted according to the transmission distance and water quality to maintain as high an SNR tolerance as possible while achieving the optimized number of bits per symbol.

TABLE II
SNR TOLERANCE AND NUMBER OF BITS PER SYMBOL FOR PAM SIGNALS

|  | Pure PAM2 | TDHP | Pure PAM4 |
|---|---|---|---|
| SNR tolerance | High | Middle | Low |
| Number of bits per symbol | 1 (Low) | 1+$p$ (Middle) | 2 (High) |

## 3. Configuration

**A. SIMULATION CONFIGURATION FOR TDHP SIGNAL**

Fig. 3 shows the simulation configuration for the optical transceiver for the non-return zero (NRZ) TDHP system under pure water conditions. The two types of modulation, the NRZ PAM 2 and PAM 4 signal ratio and power ratio are configured using the $p$ and $q$ parameters.

On the transmitter side, the TDHP signal is generated. The TDHP symbol is modulated at the mapper. The TDHP signal has low power, so for the LD to transmit the signal the driver amplifier (DA) increases the TDHP signal power according to the LD transmission power configuration. Subsequently, the TDHP signal is transferred to the LD. The optically modulated TDHP signal is transmitted to the receiver via the pure water channel.

On the receiver side, the PD detects the TDHP signal. The received TDHP signal power is low due to the underwater transmission, so a transimpedance amplifier (TIA) increases the TDHP signal power. Subsequently, the received TDHP signal is demodulated at the demapper.

Then, the TDHP bit sequence is used to calculate the SNR with respect to the BER. Based on the results, an SNR-BER graph is generated. A lower BER means high TDHP performance, while a higher BER means low TDHP performance. The TDHP performance improves by adjusting the $p$ and $q$ parameters.

The simulation parameters are shown in Table III. There are two types of generated signals: PAM 2 and PAM 4. We combine these signals to create a TDHP signal. The $p$ and $q$ parameters control the signal ratio and power ratio of the PAM 2 and PAM 4 signals to optimize the TDHP signal. The $p$ parameter controls the NRZ and PAM 4 signal ratio. By increasing $p$, the PAM 4 signal ratio increases, and the PAM 2 signal ratio decreases. Overall, the TDHP signal data rate increases. The $q$ parameter controls the power ratio between the PAM 2 and PAM 4 signals. Increasing $q$ increases the PAM 4 signal power while the PAM 2 signal power decreases simultaneously.

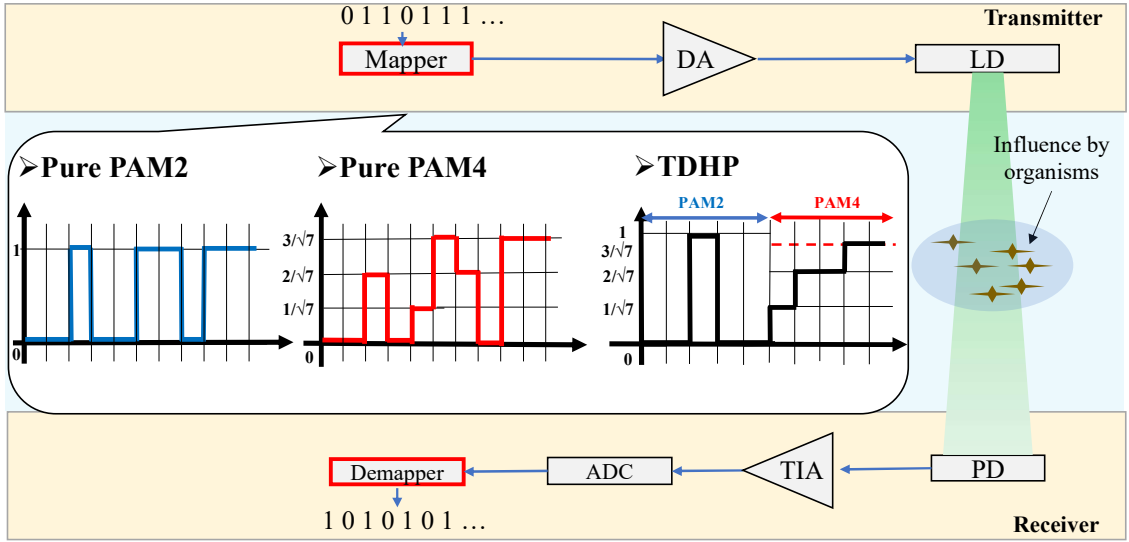

FIGURE 3.   Block diagram of simulation.

TABLE III
PARAMETERS FOR NUMERICAL CALCULATION

| Parameter | Symbol | Unit | Value |
|---|---|---|---|
| Transmission power | $P_t$ | Watt | 0.5 |
| Noise power | $P_n$ | Watt | $2 \times 10^{-6}$ |
| Beam attenuation coefficient | $K$ | 1/Meter | Red light (650 [nm]): $3 \times 10^{-1}$ <br> Green light (550 [nm]): $7 \times 10^{-2}$ <br> Blue light (450 [nm]): $2 \times 10^{-2}$ |
| Receiver aperture diameter | $D$ | Meter | 0.2 |
| Focal length | $F$ | Meter | 0.6 |
| Angle misalignment between the optical axis of the receiver and the line-of-sight between Tx and Rx | $\varphi$ | Degree | 10 |
| Half-angle Tx beamwidth | $\theta$ | Degree | 10 |
| Receiver field-of-view | FOV | Degree | 10 |
| Noise equivalent power | NEP | Watt / $\sqrt{Hz}$ | $0.4 \times 10^{-12}$ |
| Transmission bandwidth | BW | Hz | $1 \times 10^9$ |

**B. CALCURATION OF MAXIMUM TRANSMISSION LENGTH**

We propose the UWOC system structure comprising an LD and PD [32]. The LD acts as the transmitter that converts the electronic signal into an optical intensity signal. Meanwhile, the PD is a receiver that converts the transmitted light signal into an electronic signal again. These two components are crucial to the UWOC system to send the TDHP signals through the underwater channel.

On the transmitter side, the ratio of PAM 2 and PAM 4 of the TDHP signal is adjusted, and then the symbol pattern of the TDHP signal is mapped at the symbol mapper. The TDHP signal is generated and then converted to an analog electrical signal in the digital-to-analog converter. Subsequently, the TDHP signal is launched into the input of the LD. Here, the TDHP signal is optically modulated to enable transmission using optical light. The transmitted TDHP signal then reaches the receiver via the underwater channel.

On the receiver side, the PD detects the TDHP signal. We note here that signal deterioration may occur during the LD-PD signal transmission. Signal deterioration is due to two main factors: transmission distance and water turbidity. In terms of water turbidity, not only water molecules but also the influence of organisms should be taken into the calculations to simulate real-world surroundings. The TDHP signal is converted into a digital signal using an analog-to-digital converter (ADC). From here, the digital TDHP signal is reconstituted into original bit sequence via demapping.

Three types of LDs, red light, green light, and blue light, are achieved by changing beam attenuation coefficient $K$. The SNR equation used in the simulation is

$$\text{SNR} = \frac{P_t D^2 \cos\varphi}{4(\tan^2\theta)P_n} \times \frac{e^{-KL_{max}}}{L_{max}^2} \qquad (7)$$

where $P_t$ is the transmission power, $D$ is the receiver aperture, $\varphi$ is the angle misalignment between the optical axis of the receiver and the LOS of the transmitter and the receiver, $K$ is the beam attenuation coefficient, $L_{max}$ is the maximum transmission distance, $\theta$ is the half-angle transmitter beamwidth, and $P_n$ is the noise power. We explain the details of parameter $\theta$ in Section II.B. For noise power $P_n$, we assume that the power of the received optical signal is so low that the contribution of shot noise to $P_n$ is negligible. Therefore, we use a constant value that constitutes thermal noise only. Aperture $D$ can be calculated using

$$D = 2F\tan(\text{FOV}) \qquad (8)$$

where FOV is the receiver field-of-view, and $F$ is the receiver focal length [33, 34]. In the simulations all parameters can be controlled. However, to obtain the value of $L_{max}$, the SNR equation must be recalculated as

$$e^{KL_{max}} \cdot L_{max}^2 = \frac{P_t D^2 \cos\varphi}{4(\tan^2\theta)P_n \text{SNR}} \qquad (9)$$

Subsequently, beam attenuation coefficient $K$, half-beamwidth $\theta$, misalignment $\varphi$, and the FOV are used as controlled parameters.

We use the LD to transmit TDHP signals because it can support a high transmission capacity of approximately 1 Gsymbol/s. In addition, LD lights are energy-efficient and yield no electromagnetic interference making the LD light a prime choice for the UWOC system. The three LD light colors of red, green, and blue are studied to determine the maximum transmission distance for each color. In terms of angle deviation, the default angle is set to 10º being the variable in the simulation.

Only simulations were performed. While the light intensity is kept constant, we measure the maximum transmission distance of red, green, and blue according to half-beamwidth $\theta$, misalignment $\varphi$, and the receiver FOV. Other remaining parameters are only for monitoring variables, so no simulations are conducted in which those parameters are varied.

# 4. Results

## A. NOISE TOLERANCE PERFORMANCE

To obtain an optimized TDHP signal, we measure and select the lowest forward error correction (FEC) limit with a threshold of BER = $3.4 \times 10^{-3}$ for TDHP for each power ratio when the signal ratio is kept constant. In this way, we can obtain the optimum TDHP signal for each $p$ and $q$ range from 0 to 1 in increments of 0.1 for each. When the PAM 4 ratio $p$ is equal to 0 or 1, the generated signal comprises pure PAM 2 and pure PAM 4 signals. Thus, the $q$ parameter is not considered.

Fig. 4 shows the FEC limit of the optimum and non-optimum TDHP signals for each PAM 4 ratio $p$. The figure shows that when PAM 4 ratio $p$ is 0 or 1, the optimum and non-optimum FEC limits are the same. However, when $p$ =0.1 to 0.9, the FEC limit of the optimum TDHP is lower than the non-optimum TDHP with a difference of approximately -2 dB. We use two methods to verify the FEC limit value: numerical calculation and Monte Carlo simulation.

As an example, Fig. 5 shows the FEC limit for each $q$ parameter when the $p$ parameter is fixed to 0.5. The results show that as $q$ increases, the FEC limit decreases until 0.6, which is the optimum FEC limit for $p = 0.5$. After the optimum point, the FEC limit increases. The calculation is performed for $q = 0.1$ to 0.9.

From these results, we can see that the difference between the optimum and non-optimum results is approximately 2 dB. For the optimum TDHP, when the PAM 4 ratio $p$ is between 0.1 to 0.9, the lowest FEC limit is selected to obtain the optimum signal. For this reason excellent performance can be preserved even as the low noise tolerance PAM 4 ratio increases.

For the TDHP signal, the SNR is controlled by the $q$ parameter, while the $p$ parameter controls the BER. From these results, we conclude that when the $p$ parameter increases, the noise tolerance properties of the TDHP signals decreases. However, by using the optimal peak $q$ parameter, the performance of the TDHP signal improves.

Figs. 6(a-d), 6(e-h), 6(i-l) show the eye diagrams for pure PAM 2, pure PAM 4, and TDHP ($p = 0.5$, $q = 0$), respectively, for SNR = 10, 20, 30, and noise-free conditions. These figures show that a higher SNR reduces the distortion of the eye diagram. They also show that increasing the SNR improves the overall performance for the pure PAM 2, pure PAM 4, and TDHP signal. When the eye diagram of TDHP when p is set to 0.5 and q = 0, the PAM 2 and PAM 4 ratios are the same. The eye diagrams for TDHP show that it has an equal amount of PAM 2 and PAM 4 eye diagrams.

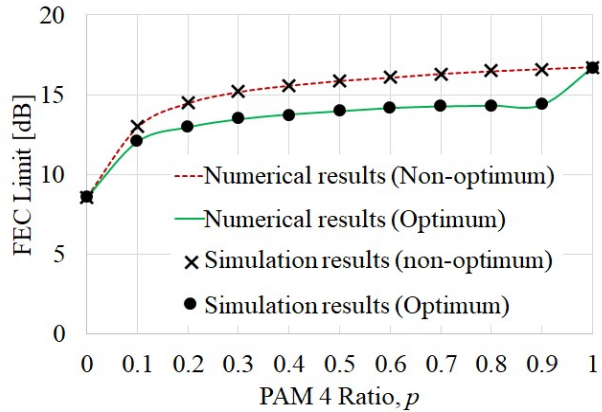

FIGURE 4.   FEC limit of non-optimum and optimum TDHP signals according to PAM 4 ratio $p$.

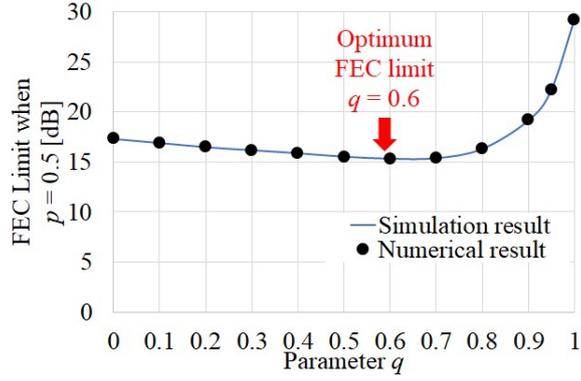

FIGURE 5. FEC limit of TDHP signal for parameter $q$ when $p = 0.5$.

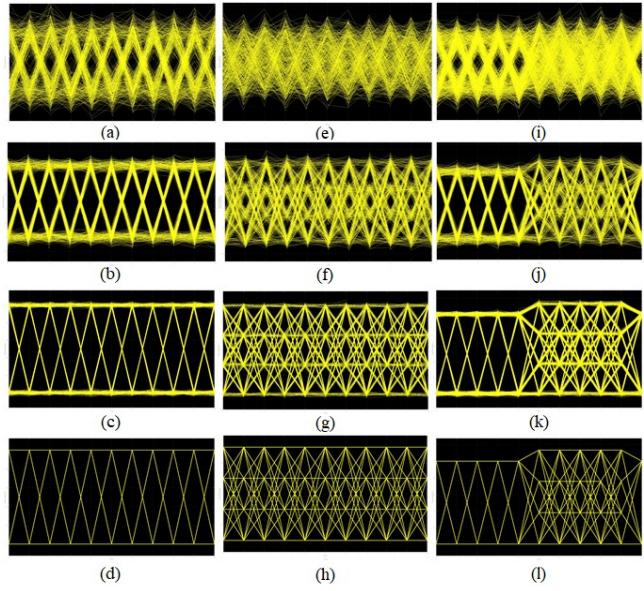

FIGURE 6. Eye diagram (a) Pure PAM 2 at SNR = 10 dB, (b) Pure PAM 2 at SNR = 20 dB, (c) Pure PAM 2 at SNR = 30 dB, (d) Pure PAM 2 at noise-free, (e) Pure PAM 4 at SNR = 10 dB, (f) Pure PAM 4 at SNR = 20 dB, (g) Pure PAM 4 at SNR = 30 dB, (h) Pure PAM 4 at noise-free, (i) TDHP ($p = 0.5$, $q = 0$) at SNR = 10 dB, (j) TDHP ($p = 0.5$, $q = 0$) at SNR = 20 dB, (k) TDHP ($p = 0.5$, $q = 0$) at SNR = 30 dB, (l) TDHP ($p = 0.5$, $q = 0$) at noise-free.

## B. SEAFLOOR TRANSMISSION PERFORMANCE

We determine the effect of PAM 4 ratio $p$ on maximum transmission distance $L_{max}$ for the red, green, and blue LDs. Fig. 7 shows the relation between PAM 4 ratio $p$ and maximum transmission distance $L_{max}$ of the optimum and non-optimum red, green, and blue lights.

The blue light has the longest $L_{max}$, followed by the green light. The red light has the shortest $L_{max}$. The red, green, and blue light wavelengths are 650 nm, 550 nm and 450 nm, respectively. The results suggest that $L_{max}$ is affected by the wavelength of the color. From the relation of the light color wavelength and $L_{max}$, we conclude that the shorter the light color wavelength, the longer $L_{max}$.

Then, we compare the non-optimum $L_{max}$ value with the optimum $L_{max}$ value to see the overall improvement in $L_{max}$. From Fig. 7, the $L_{max}$ improvement in the red, blue, and green LDs are approximately 1 m, 2.5 m, and 5 m respectively. The blue light exhibits the best improvements, followed by the green light. The red light exhibits the lowest improvements. These results suggest that the ranges affect the total distance. We conclude that the wider the difference in the highest and lowest $L_{max}$ values, the easier it is for $L_{max}$ to deteriorate.

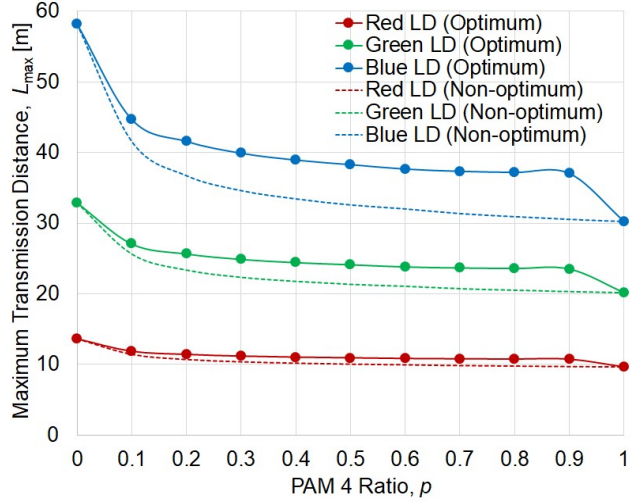

FIGURE 7. Comparison of PAM 4 ratio $p$ and maximum transmission distance $L_{max}$.

## 5. Discussion

**A. EFFECT OF HALF-ANGLE TRANSMISSION BEAMWIDTH**

In this sub-section, we evaluate the effects of half-beamwidth $\theta$ towards the overall transmission distance for the optimum $q$ value. Fig. 8 shows the effect of half-beamwidth $\theta$ on maximum transmission distance $L_{max}$ using the red LD. When comparing $L_{max}$ values for the optimum and non-optimum transmissions, the improvements are the lowest at approximately 1 m. A moderate improvement is observed from 14 to 10 m. As $p$ increases, $L_{max}$ decreases, with the lowest $L_{max}$ value occurring when $\theta = 30º$, ranging from 8.5 to 5.8 m. Fig. 9 shows the effect of half-beamwidth $\theta$ on maximum transmission distance $L_{max}$ using the green LD. When comparing $L_{max}$ values for the optimum and non-optimum transmissions, the best improvements occur at approximately 2 m. When the initial half-beamwidth $\theta = 10º$ is used, the green LD $L_{max}$ value is from 33 to 20 m. As $p$ increases, $L_{max}$ decreases, with the lowest $L_{max}$ value occurring when $\theta = 30º$, ranging from 17.5 to 9 m. Fig. 10 shows the effect of half-beamwidth $\theta$ on maximum transmission distance $L_{max}$ using the blue LD. When comparing the $L_{max}$ values for the optimum and non-optimum transmissions, the highest improvements occur at approximately 4 m. When the initial half-beamwidth $\theta = 10º$ is used, the blue LD $L_{max}$ value is from 58 to 30 m. As $p$ increases, $L_{max}$ decreases, with the lowest $L_{max}$ vale occurring when $\theta = 30º$, ranging from 25 to 11 m.

The results show that beamwidth $\theta$ significantly affects $L_{max}$. As half-beamwidth $\theta$ increases, $L_{max}$ decreases. The results indicate that to improve $L_{max}$, beamwidth $\theta$ must be narrowed. As transmitter half-beamwidth $\theta$ increases, the beam intensity decreases leading to a lower $L_{max}$ value.

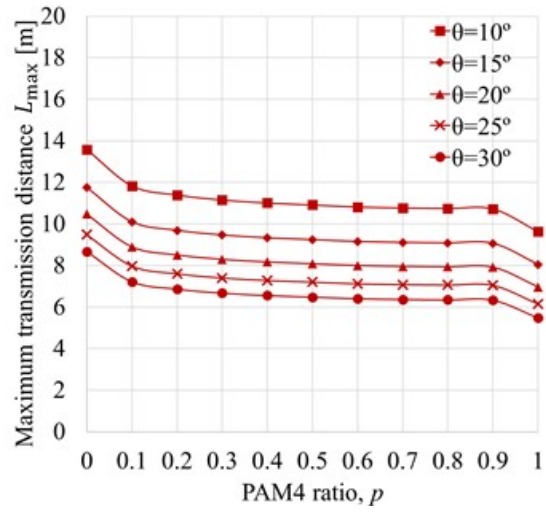

FIGURE 8. Effect of half-beamwidth $\theta$ on maximum transmission distance $L_{max}$ for red LD.

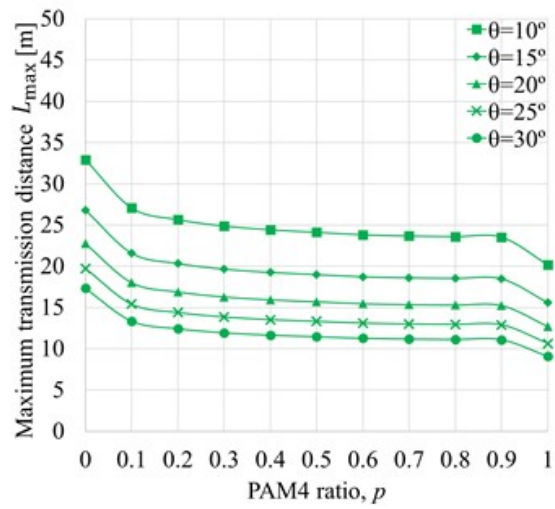

FIGURE 9. Effect on half-beamwidth $\theta$ on maximum transmission distance $L_{max}$ for green LD.

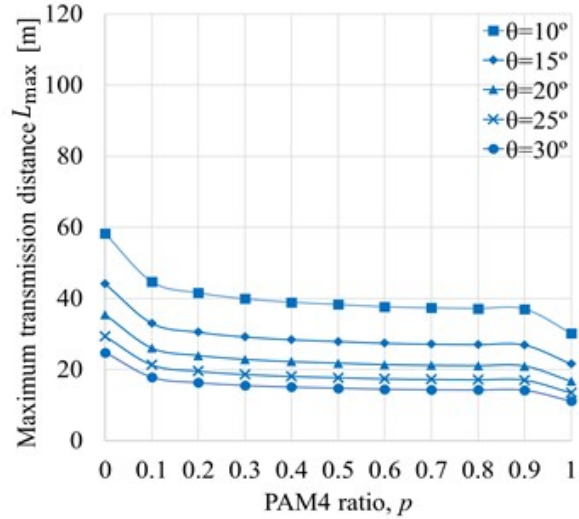

FIGURE 10. Effect on half-beamwidth $\theta$ on maximum transmission distance $L_{max}$ for blue LD.

**B. ANGLE MISALIGNMENT BETWEEN RECEIVER OPTICAL AXIS AND LOS**
A change in misalignment $\varphi$ could affect the transmission distance. In the case of the optimum $q$ and $\varphi = 0º$, the Tx-Rx LOS and the receiver optical axis are equal. Therefore, the maximum transmission distance is the highest.

The effects of the receiver FOV angle towards $L_{max}$ are evaluated. The $L_{max}$ value ranges from 5º to 25º in increments of 5º for each measurement. For this simulation, the value of $\varphi$ is 5º, 10º, 15º, 20º, and 25º.

Fig. 11 shows the effect of misalignment $\varphi$ on maximum transmission distance $L_{max}$ using the red LD. When comparing the $L_{max}$ values for the optimum and non-optimum transmissions, the lowest improvements occur at approximately 0.8 m. When the initial misalignment $\varphi = 10º$ is used, the red LD $L_{max}$ value is 11 to 14 m. As $p$ increases, $L_{max}$ decreases with the lowest $L_{max}$ value occurring when $\varphi = 50º$ ranging from 9 to 13 m.

Fig. 12 shows the effect of misalignment $\varphi$ on maximum transmission distance $L_{max}$ using the green LD. When comparing the $L_{max}$ values for the optimum and non-optimum transmissions, the best improvements occur at approximately 2.6 m. When the initial value of misalignment $\varphi = 10º$ is used, the green LD $L_{max}$ value is 33 to 20 m. As $p$ increases, $L_{max}$ decreases with the lowest $L_{max}$ value occurring when $\varphi = 50º$ ranging from 18 to 30 m.

Fig. 13 shows the effect of misalignment $\varphi$ on maximum transmission distance $L_{max}$ using the blue LD. When comparing the $L_{max}$ values for the optimum and non-optimum transmissions, the best improvements occur at approximately 5 m. When the initial misalignment $\varphi = 10º$ is used, the blue LD $L_{max}$ value is 31 to 60 m. As $p$ increases, $L_{max}$ decreases, with the lowest $L_{max}$ value occurring when $\varphi = 50º$ ranging from 27 to 52 m.

From these results, we observe that misalignment $\varphi$ has little to no effect on $L_{max}$. As misalignment $\varphi$ increases, $L_{max}$ decreases by a small amount. Thus, narrowing the misalignment of $\varphi$ could improve $L_{max}$, but because the improvement is not significant, we will not consider improving misalignment $\varphi$.

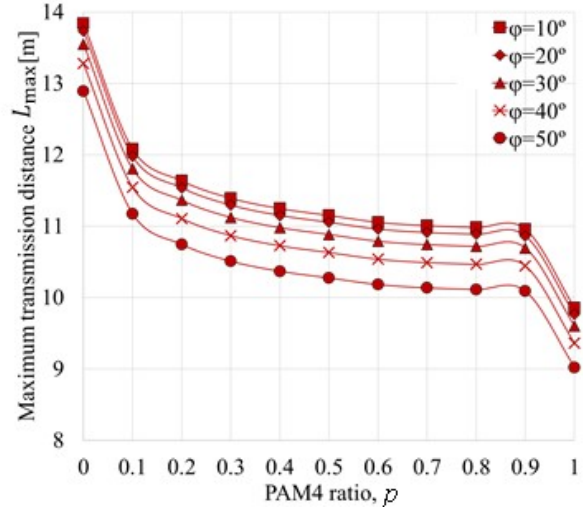

FIGURE 11. Effect of misalignment $\varphi$ on maximum transmission distance $L_{max}$ for red LD.

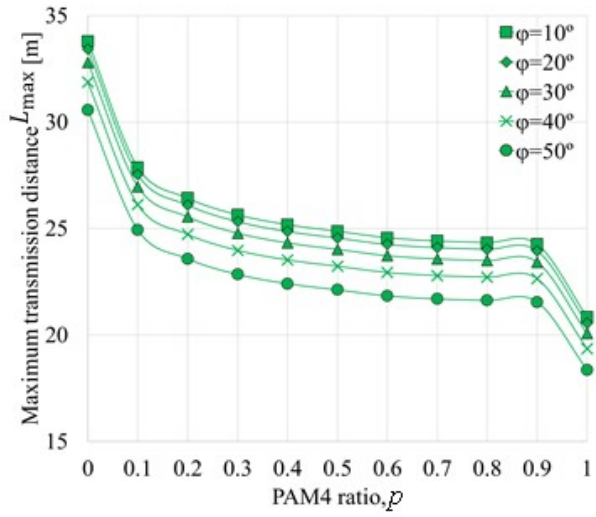

FIGURE 12. Effect of misalignment $\varphi$ on maximum transmission distance $L_{max}$ for green LD.

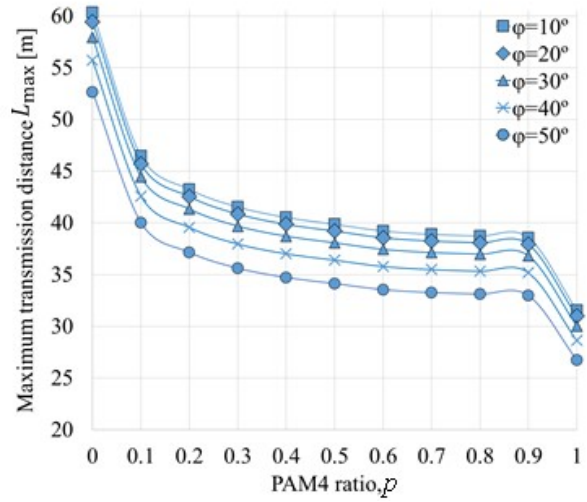

FIGURE 13. Effect of misalignment $\varphi$ on maximum transmission distance $L_{max}$ for blue LD.

## C. EFFECT OF FOV

A change in the receiver-side FOV could affect the performance of the receiver for the optimum $q$ value. The FOV of an optical system is the maximum angular size where the PD can receive the optical signals from an LD. A receiver with the same size as the receiver aperture diameter is preferred because it is easier to receive the incoming optical signals even when misalignment occurs.

The effects of the receiver FOV towards $L_{max}$ are evaluated. The $L_{max}$ values range from 5º to 25º in increments of 5º for each measurement. In this simulation, the value of φ is 5º, 10º, 15º, 20º, and 25º.

Fig. 14 shows the effect of the receiver FOV on maximum transmission distance $L_{max}$ using the red LD. When comparing the $L_{max}$ values for the optimum and non-optimum transmissions, the lowest improvements occur at approximately 0.8 m. When the initial receiver FOV = 5º is used, the red LD $L_{max}$ is from 8 to 11 m. As $p$ increases, $L_{max}$ decreases with the highest $L_{max}$ value occurring when receiver FOV = 25º ranging from 14 to 18 m.

Fig. 15 shows the effect of the receiver FOV on maximum transmission distance $L_{max}$ using the green LD. When comparing the $L_{max}$ values for the optimum and non-optimum transmissions, the best improvements occur at approximately 1.7 m. When the initial receiver FOV = 10º is used, the green LD $L_{max}$ value is from 13 to 23 m. As $p$ increases, $L_{max}$ decreases with the highest $L_{max}$ receiver FOV = 25º ranging from 13 to 24 m.

Fig. 16 shows the effect of the receiver FOV on maximum transmission distance $L_{max}$ using the blue LD. When comparing the $L_{max}$ values for the optimum and non-optimum transmissions, the best improvements occur at approximately 6 m. When the initial receiver FOV = 10º is used, the blue LD $L_{max}$ value is from 18 to 37 m. As $p$ increases, $L_{max}$ decreases with the highest $L_{max}$ value occurring when FOV = 25º ranging from 18 to 37 m.

From these results, we observe that the receiver-side FOV significantly affects $L_{max}$. As the receiver FOV increases, $L_{max}$ increases significantly. Widening the receiver-side FOV could improve the performance of $L_{max}$. As the receiver FOV increases, the easier it is for the receiver to receive the transmitted signal.

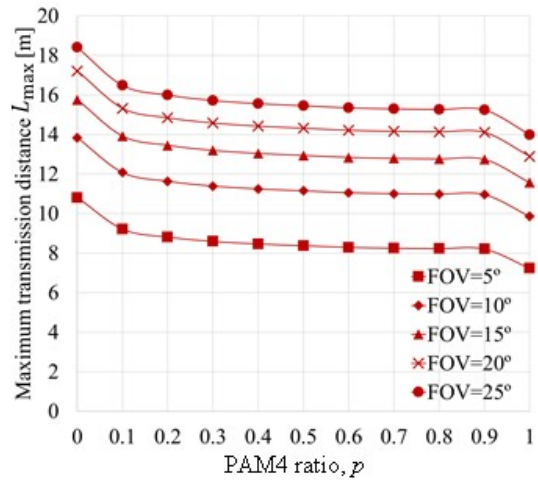

FIGURE 14. Effect of receiver FOV on maximum transmission distance $L_{max}$ for red LD.

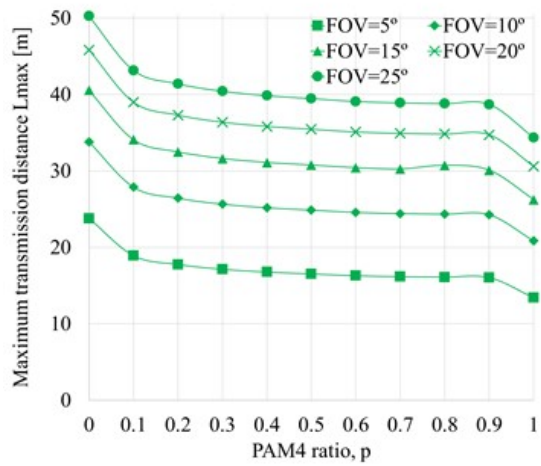

FIGURE 15. Effect of receiver FOV on maximum transmission distance $L_{max}$ for green LD.

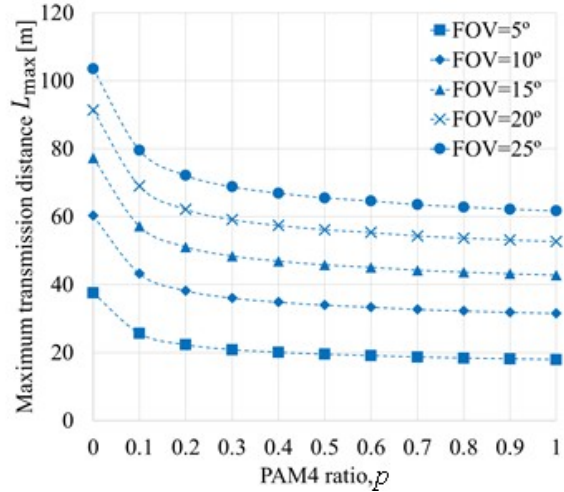

FIGURE 16. Effect of receiver FOV on maximum transmission distance $L_{max}$ for blue LD.

## 6. Conclusions

We concluded that the optimization of the TDHP signal significantly improves the overall performance from the following steps. For selecting the LD, green is chosen because of its optimum wavelength that reduces the absorption/scattering effect of the transmitted signal. Finally, by narrowing half-beamwidth $\theta$ and improving the receiver FOV, the overall performance of the UWOC system is improved.

When comparing the research objectives with the simulation results obtained, it is possible to say that the objectives of this research were achieved. The remaining objective of testing the UWOC system performance in a non-controlled environment, such as in the ocean or dirty water, was not explored due to complications with the simulation configuration and time constraints, leading to a reduced test schedule.